\documentclass{iaus}
\usepackage{graphicx}

\title[JD 11.~~Revision of Star-Formation Measures] 
{Revision of Star-Formation Measures}

\author[Claus Leitherer]   
{Claus Leitherer}

\affiliation{Space Telescope Science Institute, 3700 San Martin Drive, Baltimore, 
MD 21218, USA \\ email: {\tt leitherer@stsci.edu} }

\pubyear{2008}
\volume{255}  
\pagerange{119--126}
\jname{Low-Metallicity Star Formation: From the First Stars to Dwarf Galaxies}
\editors{L.K. Hunt, S. Madden \& R. Schneider, eds.}
\begin{document}

\maketitle

\begin{abstract}

Rotation plays a major role in the evolution of massive stars. A revised grid of stellar
evolutionary tracks accounting for rotation has recently been released by the Geneva group and
implemented into the Starburst99 evolutionary synthesis code. Massive stars are predicted
to be hotter and more luminous than previously thought, and the spectral energy distributions of
young populations mirror this trend. The hydrogen ionizing continuum in particular increases by
a factor of up to 3 in the presence of rotating massive stars. The effects of rotation generally increase towards shorter wavelengths and with decreasing metallicity. Revised
relations between star-formation rates and monochromatic luminosities for the new stellar 
models are presented. 

\end{abstract}

\firstsection 
\section{Background}

The star-formation rates of galaxies are commonly determined from the integrated light emitted at
a certain wavelengths, such as the $V$ band. Comparison with theoretically predicted  mass-to-light 
($M/L$) ratios can in principle provide the total stellar mass and, in combination with an 
appropriate timescale, the star-formation rate. This fundamental methodology goes back to Tinsley's
(1980) pioneering work. Although the reasoning is immediately intuitive, two major challenges need to
be tackled.

The first challenge is the a priori unknown stellar initial 
mass function (IMF). Observed values of $M/L_{\rm V}$ in galaxy centers are around 5~--~100. At the same time we know
that the main contributors to the galaxy light are upper main-sequence (MS) and evolved low-mass stars.
These stars have $M/L_{\rm V} \approx 1$. To put these numbers into perspective, the average $M/L_{\rm V}$ of
all known stars within 20~pc of the Sun is about 1~--~2 (Faber \& Gallagher 1979), and an early-type MS star has $M/L_{\rm V} \approx 10^{-2}$.
Since the mass-to-light ratio in galaxy centers is not too sensitive to dark matter (at least for disk galaxies;
cf. E. Brinks' talk at this conference), the apparent discrepancy suggests that most of the stellar mass
is hidden from view because most stars have lower luminosity, and therefore have
lower mass than indicated by the spectrum. 
For the purpose of this paper I will assume we can correct for this effect by assuming a known, universal 
IMF (Kroupa 2007 and this conference).

The second challenge involves the $M/L$ of individual stars of all masses. At the high-mass end, this quantity
is not accessible to direct measurements and can only be predicted by stellar evolution models: masses are poorly
known because of the scarcity of very massive binaries with mass determinations, and luminosities are
elusive because most of the stellar light is emitted in the ionizing ultraviolet. The purpose of
this paper is to discuss how the latest generation of stellar evolution models including rotation differs from
its predecessor, and how these new models affect the predictions of the evolutionary synthesis code Starburst99
(Leitherer et al. 1999; V\'azquez \& Leitherer 2005). Some of the results presented here can be found in 
V\'azquez et al. (2007).

\section{Stellar Evolution Models with Rotation}

Until the late 1990's the evolution of massive stars was thought to be determined by the chemical composition,
stellar mass, and mass-loss rate, plus atomic physics and some secondary adjustable parameters. The resulting
model grid led to reasonable agreement both with observations of individual stars and of stellar populations. Subsequently it was recognized that stellar rotation can play a key role in the
evolution of massive stars (Maeder \& Meynet 2000). Evidence of anomalous stellar surface abundances on the MS, 
lifetimes of certain evolutionary phases, and revised lower mass-loss rates support the concept of rotation.
Rotation modifies the hydrostatic structure, induces 
additional mixing and affects the stellar mass loss (cf. R. Hirschi's conference contribution).  

In Fig.~1 I illustrate the effects of rotation on the evolution of a 60~$M_\odot$ star. Rotation leads to
generally higher luminosities and higher effective temperatures for massive stars. This is the result of
the larger convective core and the lower surface opacity in the presence of rotation. (Recall that hydrogen is
the major opacity source and any decrease of its relative abundance by mixing lowers the opacity and therefore increases
the temperature.) Fig.~1 suggests a significant luminosity increase of a 60~$M_\odot$ star even on the 
MS. This trends is present in all massive stars down to $\sim$20~$M_\odot$, depending on metallicity. The
lower the metallicity, the more important the influence of rotation, which ultimately becomes the dominant
evolution driver for metal-free stars (Hirschi et al. 2008).

\begin{figure}
\begin{center}
 \includegraphics[width=4in]{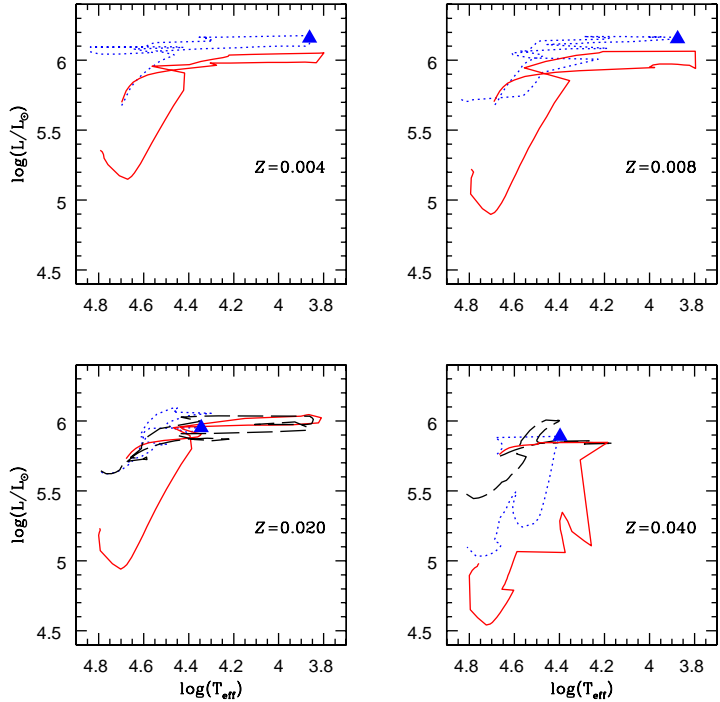} 
 \caption{Comparison of the evolution of a 60~$M_\odot$ star without (solid) and with (dotted) rotation for 
four different metal abundances. $v_{\rm rot} = 300$~km~s$^{-1}$. The line denotes an exploratory model
with $v_{\rm rot} = 0$ and all other parameters identical to the rotating model.
See V\'azquez et al. (2007) for additional details.}
\end{center}
\end{figure}

\section{Revised Population Models}

V\'azquez et al. (2007) implemented the full grid of rotating evolution models into Starburst99.
Stellar atmospheres and/or empirical spectral libraries were attached for each mass and at each
time step. The atmospheres that were used for hot stars are those published by Smith, Norris, \& Crowther (2002).
For an assumed IMF one can then compute the full spectral energy distribution (SED) and its evolution with
time. All models quoted here are for a Salpeter IMF with mass cut-offs at 1 and 100~$M_\odot$. The Hawaii
group is independently using these SED's as input for photo-ionization modeling with the Mappings code (E. Levesque, these proceedings).
 
\begin{figure}
\begin{center}
 \includegraphics[width=3.8in]{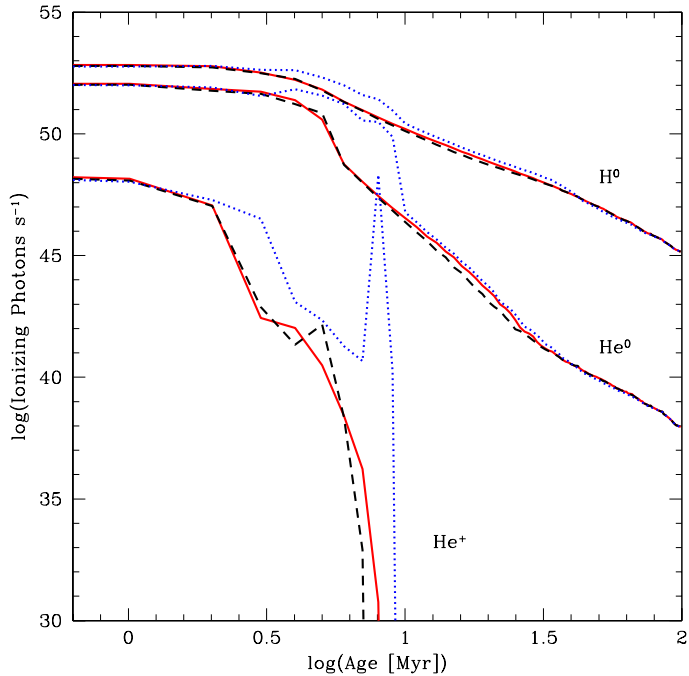} 
 \caption{Number of photons in the H$^0$, He$^0$, and He$^+$ continuum (solar composition). Line types as in Fig.~1.}
\end{center}
\end{figure}

The most dramatic changes with respect to prior models occur at the short-wavelength end of the SED. The ionizing
luminosities for a singular burst with mass $10^6$~$M_\odot$ are shown in Fig.~2. Since the most massive stars
are more luminous and hotter, their ionizing luminosities increase during O-star dominated phases (2~--~10~Myr).
The increase reaches a factor of 3 in the hydrogen ionizing continuum and several orders of magnitude in 
the neutral and ionized helium continua. The predictions for the latter need careful scrutiny, as the photon
escape fraction crucially depends on the interplay between the stellar parameters supplied by the evolution
models and the radiation-hydrodynamics of the atmospheres. In contrast, the escape of the hydrogen ionizing
photons has little dependence on the particulars of the atmospheres and consequently is a relatively safe 
prediction.

Luminosities for selected wavelengths and passbands are reproduced in Fig.~3. 
In addition to the previously discussed changes when O stars dominate, the figure suggests significant revisions
at epochs when red supergiants are present (10~--~20~Myr). The new evolution models with rotation predict an
enhanced red supergiant phase which becomes noticeable, e.g., in the higher $K$ band luminosity. The fact that
all curves converge after $\sim$50~Myr is an artifact: the models with rotation only reach down to a mass of 9~$M_\odot$, and
the traditional tracks were used at masses below that value. However, the effects of rotation are expected to be
small at these lower masses.

\begin{figure}
\begin{center}
 \includegraphics[width=4.3in]{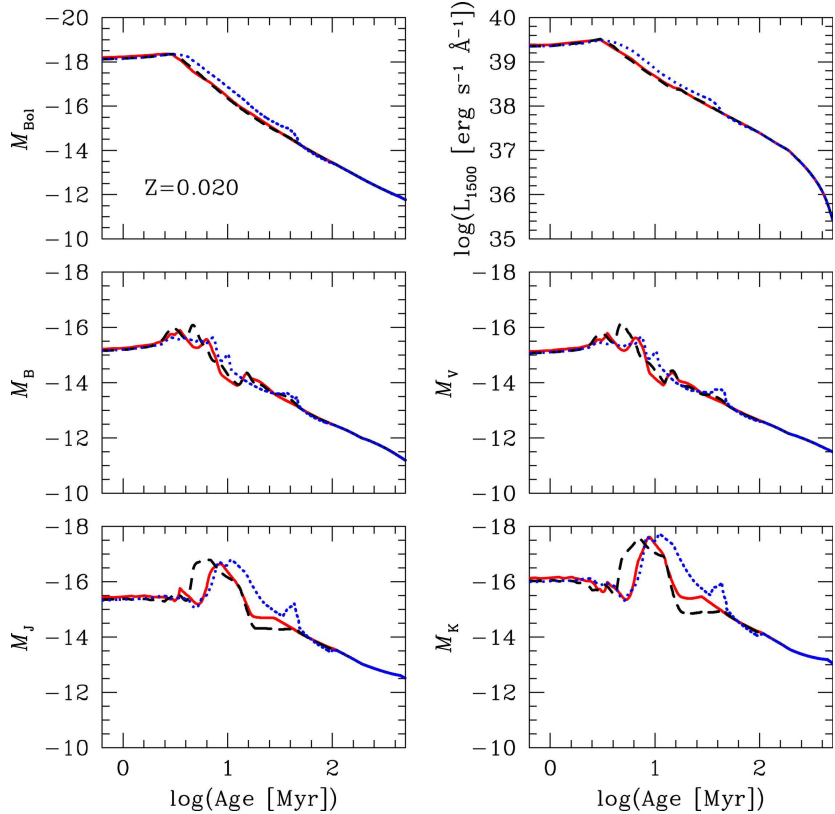} 
 \caption{$M_{\rm Bol}$, $L_{\rm 1500}$, $M_{\rm B}$, $M_{\rm V}$, $M_{\rm J}$, and $M_{\rm K}$ vs. time. The bands most affected by the new models are those in the ultraviolet and infrared. Solar chemical composition. Line types as in Fig.~1.}
\end{center}
\end{figure}

 \section{Implications for Star-Formation Indicators}

The results discussed so far apply to stellar populations forming quasi-instantaneously. Choosing an
instantaneous population makes it easier to identify physical processes in the SED since a particular epoch in time is
usually associated with a specific stellar mass interval. While singular bursts are a good approximation for
the star-formation history of, e.g., a stellar cluster, galaxies are better described by a star-formation 
equilibrium when stellar birth and death rates are identical. In this case one can derive relations between the
star-formation rate and monochromatic luminosity independent of age.

Star-formation rates as a function of luminosity for several strategic wavelengths were determined for steady-state
populations of age 100~Myr. At that epoch massive stars have reached an equilibrium for all wavelengths considered here. The IMF is the same as before. The new relations for stellar models with rotation having solar chemical composition are: 
\begin{eqnarray}
\label{eq:LLyc}
SFR~[M_{\odot}~\mathrm{yr}^{-1}] = 
      3.55 \times 10^{-54} N_{\rm{LyC}}~[\mathrm{s}^{-1}]\\
\label{eq:L1500}
SFR~[M_{\odot}~\mathrm{yr}^{-1}] = 
      3.39 \times 10^{-41} L_{\rm{1500}}~[\mathrm{erg~s}^{-1}~\mathrm{\AA}^{-1}]\\
\label{eq:LV}
SFR~[M_{\odot}~\mathrm{yr}^{-1}] =
      6.31 \times 10^{-40} L_{\rm{V}}~[\mathrm{erg~s}^{-1}~\mathrm{\AA}^{-1}]\\
\label{eq:LIR}
SFR~[M_{\odot}~\mathrm{yr}^{-1}] = 
      1.48 \times 10^{-44} L_{\mathrm{FIR}}~[\mathrm{erg~s}^{-1}].
\end{eqnarray}
For comparison, if eqs.~(\ref{eq:LLyc}), (\ref{eq:L1500}),(\ref{eq:LV}), and (\ref{eq:LIR}) were
derived with the previous tracks (as currently implemented in Starburst99), the conversion factors between
luminosity and star-formation rate would be $4.42 \times 10^{-54}$, $4.07 \times 10^{-41}$, $6.76 \times 10^{-40}$, and
$1.78 \times 10^{-44}$, respectively. The revised relations lead to somewhat lower star-formation rates when applied
to the commonly used star-formation measures. The largest effect is for the ionizing photon flux, which can be determined, e.g., from the H$\alpha$ luminosity. The new rates will be about 25\% lower for the same H$\alpha$ luminosity. In practice,  this
decrease is hardly significant because other systematic uncertainties, such as the IMF scaling, are more important. 

For lower metallicities, the $M/L$ of rotating stars becomes even lower, and this trend is reflected in the $M/L$ of the populations. Consequently the conversion coefficients in eqs.~(\ref{eq:LLyc}), (\ref{eq:L1500}),(\ref{eq:LV}), and (\ref{eq:LIR}) for 20\% solar composition become $2.77 \times 10^{-54}$, $3.31 \times 10^{-41}$, $5.50 \times 10^{-40}$, and
$1.43 \times 10^{-44}$, respectively. The difference between the new conversion at 20\% and the previous conversion at solar composition for the ionizing luminosity reaches almost a factor of 2, which is clearly non-negligible. 

To summarize, we find noticeable changes in the theoretically predicted $M/L$ ratios of stellar populations computed with the new grid of stellar evolutionary tracks with rotation. Whenever hot, massive stars contribute to the SED, the revised
luminosities are higher and the spectrum is harder. The effects are subtle at optical and infrared wavelengths but significant in the ultraviolet. Single stellar populations with ages of several Myr are predicted to have ionizing fluxes that are higher by a factor of up to 3. Steady-state populations are less affected because of the diluting effect of ongoing star formation. Nevertheless, the conversion factor between H$\alpha$ luminosity and star-formation rate may change by 25\% or more. 

A prudent Starburst99 user may want to take the new calibrations with care. While there is general consensus that the
new evolution models with rotation are a quantum leap over their predecessors, these tracks are still in an exploratory stage and further testing is needed. Ultimately, the new grid and the corresponding revision of the star-formation measures will become the default in Starburst99. It is frustrating from the perspective of the evolutionary synthesis modeler that stellar rotation introduces a new free parameter that reduces some of the deterministic concepts of the previous model generation.

\end{document}